\documentclass[twocolumn,showpacs,preprintnumbers,amsmath,amssymb,prb,superscriptaddress]{revtex4-1}
\usepackage{mathrsfs}
\usepackage{graphicx}
\usepackage{dcolumn}
\usepackage{bm}
\usepackage{amsmath}
\usepackage{amsfonts}
\usepackage{color}

\begin{document}

\title{Chiral topological superconductor and half-integer conductance plateau from quantum anomalous Hall plateau transition}
\author{Jing Wang}
\affiliation{Department of Physics, McCullough Building, Stanford University, Stanford, California 94305-4045, USA}
\affiliation{Stanford Institute for Materials and Energy Sciences, SLAC National Accelerator Laboratory, Menlo Park, California 94025, USA}
\author{Quan Zhou}
\affiliation{Department of Physics, McCullough Building, Stanford University, Stanford, California 94305-4045, USA}
\author{Biao Lian}
\affiliation{Department of Physics, McCullough Building, Stanford University, Stanford, California 94305-4045, USA}
\author{Shou-Cheng Zhang}
\affiliation{Department of Physics, McCullough Building, Stanford University, Stanford, California 94305-4045, USA}
\affiliation{Stanford Institute for Materials and Energy Sciences, SLAC National Accelerator Laboratory, Menlo Park, California 94025, USA}

\begin{abstract}
We propose to realize a two-dimensional chiral topological superconducting (TSC) state from the quantum anomalous Hall plateau transition in a magnetic topological insulator thin film through the proximity effect to a conventional $s$-wave superconductor. This state has a full pairing gap in the bulk and a single chiral Majorana mode at the edge. The optimal condition for realizing such chiral TSC is to have inequivalent superconducting pairing amplitudes on top and bottom surfaces of the doped magnetic topological insulator. We further propose several transport experiments to detect the chiral TSC. One unique signature is that the conductance will be quantized into a half-integer plateau at the coercive field in this hybrid system. In particular, with the point contact formed by a superconducting junction, the conductance oscillates between $e^2/2h$ and $e^2/h$ with the frequency determined by the voltage across the junction. We close by discussing the feasibility of these experimental proposals.
\end{abstract}

\date{\today}


\maketitle

\section{Introduction}

The search for topological states of matter has become a central focus in condensed matter physics. Chiral topological superconductors (TSC) in two-dimensions (2D) with an odd-integer Chern number are predicted to host a Majorana zero mode in the vortex core, which obeys non-Abelian statistics~\cite{read2000,ivanov2001} and has potential applications in topological quantum computation~\cite{nayak2008}. A chiral TSC with Chern number $\mathcal{N}$ breaks time-reversal symmetry, and has a full pairing bulk gap and $\mathcal{N}$ topologically protected gapless chiral Majorana edge modes (CMEMs), which can be viewed as a superconducting analogy of the quantum Hall (QH) state~\cite{volovik1988,qi2009,schnyder2008}. As a minimal topological state in 2D, the $\mathcal{N}=1$ chiral TSC is of particular interest, as its edge state has only half the degrees of freedom of the QH state with Chern number $\mathcal{C}=1$. Intensive efforts have been made to search for the chiral TSC in 2D~\cite{fu2008,fu2009a,sato2009,sau2010,alicea2010,qi2010b,ojanen2014,li2015,mackenzie2003,raghu2010,wangqh2013}, however, it has not yet been confirmed in experiments.

In principle, a QH state with Chern number $\mathcal{C}$ in proximity with an $s$-wave superconductor (SC) can be naturally viewed as a chiral TSC with even number $\mathcal{N}=2\mathcal{C}$ CMEMs. Therefore, it is theoretically possible to realize a chiral TSC with odd number of CMEMs near a QH plateau transition~\cite{qi2010b}. However, the strong magnetic field required in a QH state will severely hinder the superconducting proximity. Instead, the quantum anomalous Hall (QAH) state has a finite Chern number $\mathcal{C}$ in the absence of an external magnetic field~\cite{thouless1982,haldane1988}, which has been theoretically predicted in magnetic topological insulators (TIs) with ferromagnetic (FM) ordering~\cite{hasan2010,qi2011,qi2008,liu2008,li2010,yu2010,wang2013a,wang2013b,wang2014b,onoda2003,biswas2010} and experimentally realized (for $\mathcal{C}=\pm1$) in both Cr-doped~\cite{chang2013b,checkelsky2014,kou2014,bestwick2015,kandala2015} and V-doped~\cite{chang2015} (Bi,Sb)$_2$Te$_3$ magnetic TI thin films. More recently, a new zero-plateau QAH state with $\mathcal{C}=0$ and the plateau transitions among $\mathcal{C}=\pm1,0$ states have been theoretically predicted~\cite{wang2014a} and experimentally observed~\cite{fengy2015,kou2015}. Without requiring a large external magnetic field, the plateau transition from the $\mathcal{C}=\pm1$ QAH to the zero-plateau $\mathcal{C}=0$ state is a unique parent system for realizing a $\mathcal{N}=\pm1$ chiral TSC.

\begin{figure}[t]
\begin{center}
\includegraphics[width=3.3in]{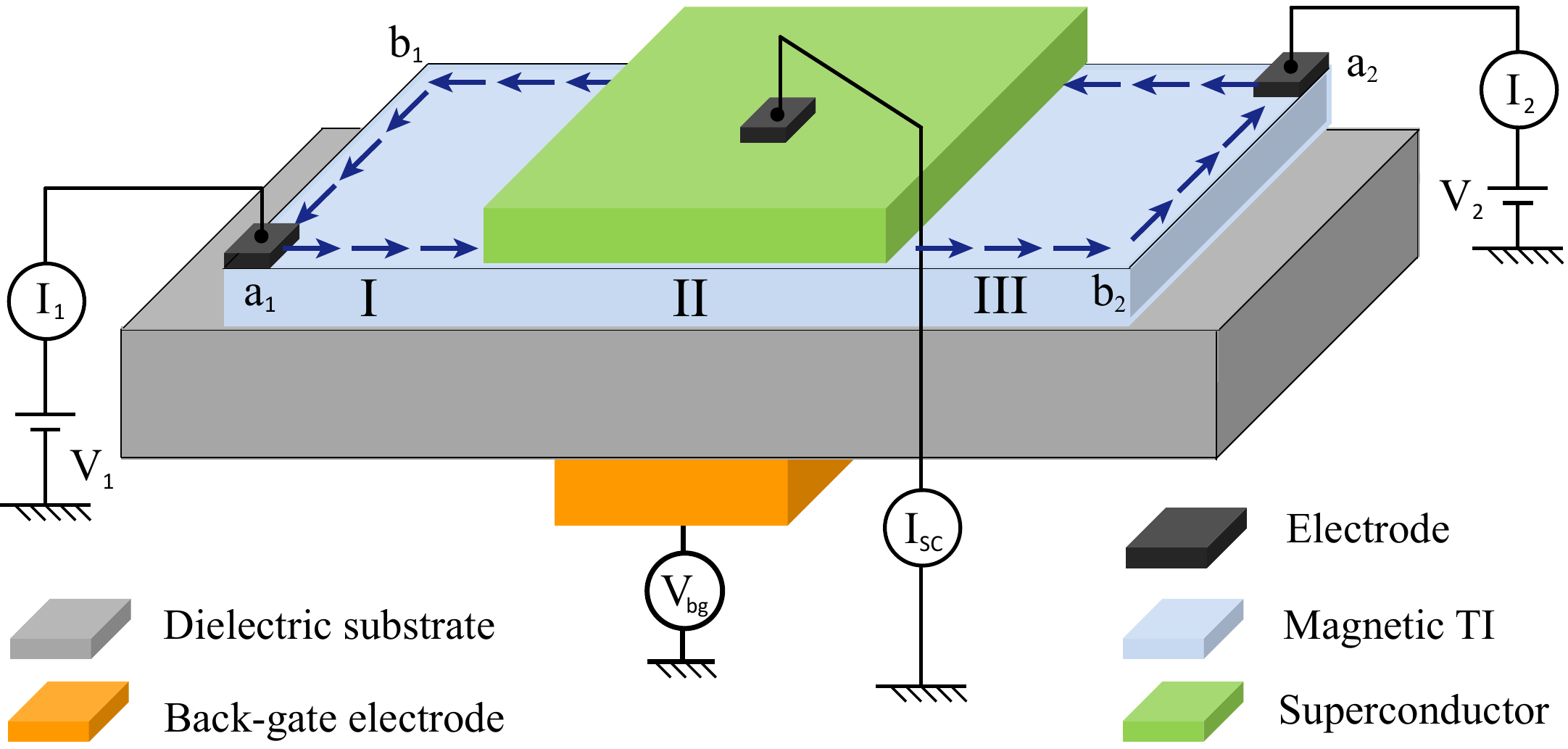}
\end{center}
\caption{(color online). The hybrid QAH-SC device. In region II, a chiral TSC state is induced through the proximity effect to an $s$-wave SC layer on top of the QAH in magnetic TI. A back-gate voltage $V_{\mathrm{bg}}$ is applied to control the Fermi level in region II. Voltages $V_1$ and $V_2$ are applied on leads 1 and 2, respectively. The SC layer is grounded through a lead in its bulk.}
\label{fig1}
\end{figure}

In this paper, we propose to realize the $\mathcal{N}=\pm1$ chiral TSC in a magnetic TI near the QAH plateau transition via the proximity effect to an $s$-wave SC. The optimal condition for realizing the chiral TSC is to have \emph{inequivalent} SC pairing amplitudes on top and bottom surfaces of the doped magnetic TI. We then propose several transport experiments to detect this chiral TSC. Generally, the conductance could be quantized into a half-integer plateau at the coercive field in this hybrid system (Fig.~\ref{fig1}), as a signature of the neutral CMEM backscattering. In particular, with a point contact formed by a SC junction (Fig.~\ref{fig4}), the conductance oscillates with a frequency determined by the voltage across the junction. Lastly, we briefly discuss the temperature dependence on the transmission of CMEM and the feasibility of these experimental proposals.

The organization of this paper is as follows. After this
introductory section, Sec. II describes the effective model
for the SC proximity effect of the QAH state in a magnetic TI thin film. Section III presents the results on the phase diagram, edge transport and experimental proposals on point contacts. Section IV
presents discussion on the feasibility of experimental realization
of chiral TSC in a magnetic TI. Section V concludes this paper. Some auxiliary materials
are relegated to appendixes.

\section{Model}

To start, we consider the SC proximity effect of the QAH state in a magnetic TI thin film with
FM order. Without the proximity effect, the low energy physics of the system only consists of the Dirac-type surface states (SS)~\cite{wang2014a}.
The 2D effective Hamiltonian is $\mathcal{H}_0=\sum_{\mathbf{k}}\psi^{\dag}_{\mathbf{k}}H_0(\mathbf{k})\psi_{\mathbf{k}}$, with $\psi_{\mathbf{k}}=(c^t_{\mathbf{k}\uparrow}, c^t_{\mathbf{k}\downarrow},c^b_{\mathbf{k}\uparrow}, c^b_{\mathbf{k}\downarrow})^T$ and
\begin{equation}\label{QAH}
H_0(\mathbf{k})=k_y\sigma_x\widetilde{\tau}_z-k_x\sigma_y\widetilde{\tau}_z+m(k)\widetilde{\tau}_x+\lambda\sigma_z,
\end{equation}
where $c_{\mathbf{k}\sigma}$ annihilates an electron of momentum $\mathbf{k}$ and spin $\sigma=\uparrow, \downarrow$, and superscripts $t$ and $b$ denote SS in the top and bottom layers, respectively. $\sigma_i$ and $\widetilde{\tau}_i$ ($i=x,y,z$) are Pauli matrices for spin and layer, respectively. $\lambda$ is the exchange field along $z$ axis induced by the FM ordering. Here $\lambda\propto\langle S\rangle$ with $\langle S\rangle$ being the mean field expectation value of the local spin, and the value of $\lambda$ can be changed during the magnetization reversal process in magnetic TIs. $m(k)=m_0+m_1(k_x^2+k_y^2)$ describes the hybridization between the top and bottom SS. The Chern number of the system is $\mathcal{C}=\lambda/|\lambda|$ for $|\lambda|>|m_0|$, and $\mathcal{C}=0$ for $|\lambda|<|m_0|$. Correspondingly, the system has $|\mathcal{C}|$ chiral edge state~\cite{wang2014a}. In proximity to an $s$-wave SC, a finite pairing amplitude is induced in the QAH system. The Bogoliubov-de Gennes (BdG) Hamiltonian becomes $\mathcal{H}_{\mathrm{BdG}}=\sum_{\mathbf{k}}\Psi^{\dag}_{\mathbf{k}}H_{\mathrm{BdG}}\Psi_{\mathbf{k}}/2$, where $\Psi_{\mathbf{k}}=[(c^t_{\mathbf{k}\uparrow}, c^t_{\mathbf{k}\downarrow}, c^b_{\mathbf{k}\uparrow}, c^b_{\mathbf{k}\downarrow}), (c^{t\dag}_{-{\mathbf{k}}\uparrow}, c^{t\dag}_{-{\mathbf{k}}\downarrow}, c^{b\dag}_{-{\mathbf{k}}\uparrow}, c^{b\dag}_{-{\mathbf{k}}\downarrow})]^T$ and
\begin{equation}\label{BdG}
\begin{aligned}
H_{\mathrm{BdG}} &= \begin{pmatrix}
H_0(\mathbf{k})-\mu & \Delta_{\mathbf{k}}\\
\Delta_{\mathbf{k}}^\dag & -H_0^*(-\mathbf{k})+\mu
\end{pmatrix},
\\
\Delta_{\mathbf{k}} &= \begin{pmatrix}
i\Delta_1\sigma_y & 0\\
0 & i\Delta_2\sigma_y
\end{pmatrix}.
\end{aligned}
\end{equation}
Here $\mu$ is chemical potential, $\Delta_1$ and $\Delta_2$ are pairing gap functions on top and bottom SS, respectively.

In a simple case for $\mu=0$ and $\Delta_1=-\Delta_2=\Delta$, a basis transformation~\cite{basis_note} decouples the BdG Hamiltonian into two models with opposite chirality, and
\begin{equation}\label{decoupled}
H_{\mathrm{BdG}} = \begin{pmatrix}
H_+(\mathbf{k}) & 0\\
0 & H_-(\mathbf{k})
\end{pmatrix},
\end{equation}
where $H_\pm(\mathbf{k})=k_y\sigma_x\mp k_x\sigma_y\varsigma_z+(m(k)\pm\lambda)\sigma_z\varsigma_z\mp\Delta\sigma_y\varsigma_y$ with $\varsigma_{x,y,z}$ the Pauli matrices in Nambu space. The topological property of $H_+$ is clearly seen by a further basis transformation into a block diagonal form:
\begin{equation}
H_+(\mathbf{k})=\begin{pmatrix}
h_+(\mathbf{k}) & 0\\
0 & -h^*_-(-\mathbf{k})
\end{pmatrix},
\end{equation}
where $h_\pm(\mathbf{k})=k_y\sigma_x-k_x\sigma_y+(m(k)+\lambda\pm|\Delta|)\sigma_z$ characterizes a $p_x \pm ip_y$ SC~\cite{read2000,fu2008}.
The BdG Chern number of $h_\pm(\mathbf{k})$ depends only on the sign of mass $m(k)+\lambda\pm|\Delta|$ at the $\Gamma$ point~\cite{wang2014a}. Therefore, the Chern number of $H_+(\mathbf{k})$ is $\mathcal{N}_+=-2$ for $|\Delta|<-m_0-\lambda$, $\mathcal{N}_+=-1$ for $|\Delta|>|m_0+\lambda|$ and $\mathcal{N}_+=0$ for $|\Delta|<m_0+\lambda$. Similarly, the Chern number of $H_-(\mathbf{k})$ is $\mathcal{N}_-=2$ for $|\Delta|<\lambda-m_0$, $\mathcal{N}_-=1$ for $|\Delta|>|m_0-\lambda|$ and $\mathcal{N}_-=0$ for $|\Delta|<m_0-\lambda$. The total Chern number of the system is then $\mathcal{N}=\mathcal{N}_++\mathcal{N}_-$.
Fig.~\ref{fig3}a shows the phase diagram of the system. The phase boundaries are determined by $\Delta\pm(m_0\pm\lambda)=0$, which reduce to the critical points $\lambda=\pm|m_0|$ between the $\mathcal{C}=\pm1$ QAH and the zero plateau normal insulator (NI) for $\Delta=0$. An infinitesimal SC gap drives the QAH phase into a $\mathcal{N}=\pm2$ TSC. More importantly, the $\mathcal{N}=\pm1$ TSC state emerges in the neighborhood of the transition between the QAH phase and NI phase.

\begin{figure}[b]
\begin{center}
\includegraphics[width=3.3in]{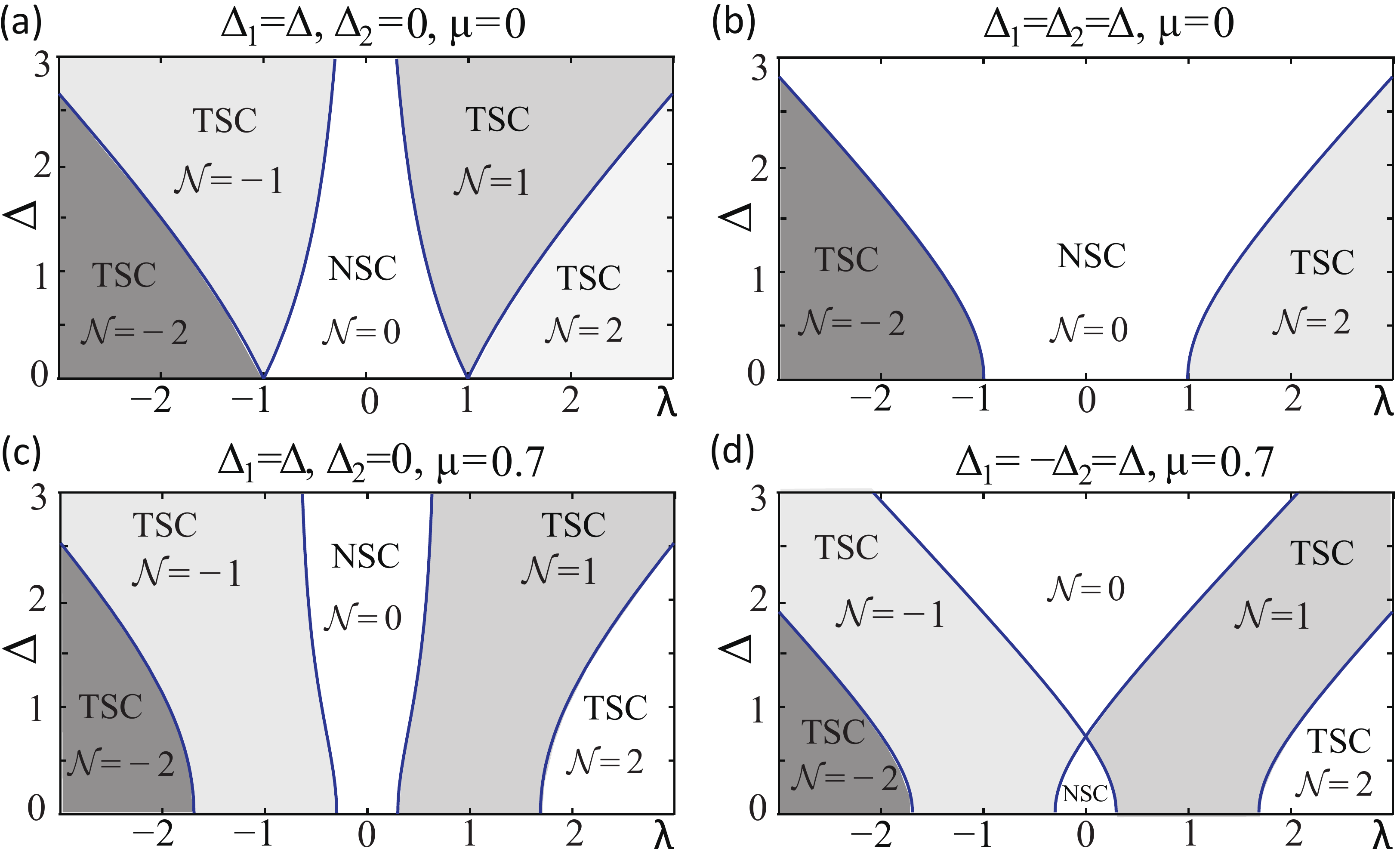}
\end{center}
\caption{Phase diagram of the QAH-SC hybrid system with typical parameters. (a) $\Delta_1=\Delta$, $\Delta_2=0$, $\mu=0$. (b) $\Delta_1=\Delta_2=\Delta$, $\mu=0$. (c) $\Delta_1=\Delta$, $\Delta_2=0$, $\mu=0.7$. (d) $\Delta_1=-\Delta_2=\Delta$, $\mu=0.7$. Here $\Delta_{1}$, $\Delta_2$, $\mu$ are in the units of $|m_0|$.}
\label{fig2}
\end{figure}

\section{Results}

\subsection{Phase diagram}

Now we turn to the optimal condition for realizing the $\mathcal{N}=\pm1$ TSC. First, consider the phase diagram for $\mu=0$ and general values of $\Delta_1$ and $\Delta_2$. The phase boundaries are determined by the bulk BdG gap closing in Eq.~(\ref{BdG}). Assuming $\Delta_2=\alpha\Delta_1$  and $\alpha$ is real, the phase boundaries are given by $\mp(1-\alpha)\Delta_1\lambda+\lambda^2=m_0^2+\alpha\Delta_1^2$, as shown in Fig.~\ref{fig2}. For $\Delta_1=\Delta_2$, the Chern number jumps directly from $\mathcal{N}=\pm2$ to $\mathcal{N}=0$, and $\mathcal{N}=\pm1$ TSC phases disappear due to accidental particle-hole symmetry in $H_0$ with $\mu=0$. As $\Delta_2$ decreases, the $\mathcal{N}=\pm1$ TSC phase space emerges and becomes the widest at $\Delta_2=0$. In particular when $\Delta_1\Delta_2<0$, a helical TSC phase with helical Majorana edge states emerges on the $\lambda=0$ line (Fig. \ref{fig3}a). The general case for complex $\alpha$ is studied in Appendix A, where the topology of phase diagram remains unchanged. Next, for the case $\mu\neq0$, which corresponds to the SC proximity effect of a \emph{doped} or \emph{electrically gated} QAH system, the proximity effect is effectively enhanced by the finite density of states at the Fermi level~\cite{qi2010b}. As shown in Fig.~\ref{fig2}, the phase space of $\mathcal{N}=\pm1$ TSC near the $\Delta=0$ axis enlarges from $\mu=0$ to $\mu\neq0$. Therefore, the optimal condition for $\mathcal{N}=\pm1$ TSC is $\mu\neq0$ and $\Delta_2=0$. This leads us to design the transport device in Fig.~\ref{fig1}. The $s$-wave SC is only grown on top of the magnetic TI in region II to ensure the proximity pairing gap of the top SS is larger than that of the bottom SS, while the Fermi level can be tuned by the back-gate. The size of the SC layer should be \emph{larger} than the back-gate electrode so that there is no metallic regions in the device. Similarly, one can also employ another device geometry by using a global back-gate and two top-gates in region I and III, to tune the Fermi levels in region I, II, and III separately.

\begin{figure}[b]
\begin{center}
\includegraphics[width=3.3in]{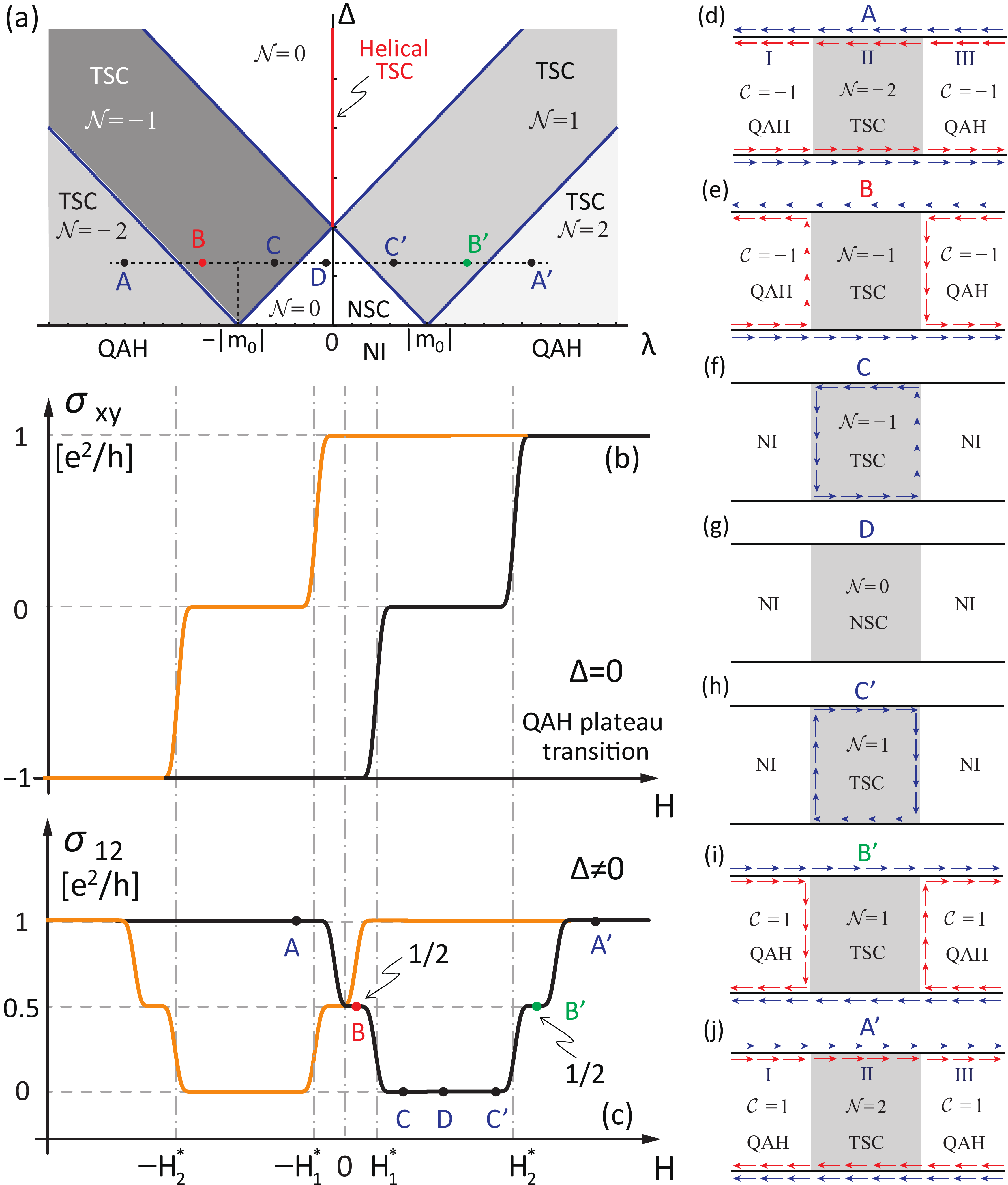}
\end{center}
\caption{(color online). (a) Phase diagram of the QAH-SC hybrid system for $\mu=0$ and $\Delta_1=-\Delta_2\equiv\Delta$.
Only $\Delta\ge0$ is shown. (b) Without SC proximity effect, the $\sigma_{xy}=-1\rightarrow0\rightarrow1$ QAH plateau transition occurs at the coercivity when the magnetization flips. (c) With SC proximity effect to region II in hybrid device Fig.~\ref{fig1}, $\sigma_{12}$ shows plateau transition $1\rightarrow1/2\rightarrow0\rightarrow1/2\rightarrow1$ in the hysteresis loop. The half-integer plateau in $\sigma_{12}$ manifests the $\mathcal{N}=1$ TSC. (d)-(j) The edge transport configuration at A, B, C, D, C$'$, B$'$ and A$'$ in (c). There is no backscattering for $\mathcal{N}=\pm2$ TSC in (d),(j), and Majorana backscattering for $\mathcal{N}=\pm1$ TSC in (e),(i). Red and blue arrows represent $(c\pm c^\dag)$ CMEMs, respectively. NSC: normal, topologically trivial SC.}
\label{fig3}
\end{figure}

\subsection{Edge transport and half-plateau}

To identify the $\mathcal{N}=1$ TSC in the QAH-SC hybrid system, one can probe the neutral Majorana nature of CMEM or trap the vortex core zero mode. Several methods have been proposed to measure the Majorana fermions~\cite{fu2009a,tanaka2009a,fu2009b,akhmerov2009,law2009,lutchyn2010,chung2011}. Here, we base our discussion on a recent proposal studying the CMEM backscattering~\cite{chung2011}. The basic setup is shown in Fig.~\ref{fig1}, consisting of a magnetic TI in proximity with a grounded top SC layer in region II and two current leads at the corners. When the magnetic domains of magnetic TI are aligned in the same direction, the magnetic TI is in a QAH state with a single chiral edge state propagating along the sample boundary. During the flipping of the magnetic domains at the coercive field, $\lambda$ decreases and the magnetic TI enters the NI with a zero-plateau in Hall conductance $\sigma_{xy}$ over a finite range of magnetic field~\cite{wang2014a,fengy2015,kou2015}, as shown in Fig.~\ref{fig3}b. Either perpendicular or in plane external magnetic field could induce such plateau transition~\cite{kou2015}.
When the SC proximity effect is sufficiently strong, the superconducting region II experiences the BdG Chern number variation $\mathcal{N}=-2\rightarrow-1\rightarrow0\rightarrow1\rightarrow2$ as $\lambda$ decreases in the hysteresis loop (dashed line in Fig.~\ref{fig3}a). Therefore, the transport setup Fig.~\ref{fig1} is a QAH/NI-TSC/NSC-QAH/NI junction. As we will discuss in details below, the edge transport features of the junction uniquely convey the topological properties of the SC in region II.

The QAH edge state can be viewed as two CMEMs since a $\mathcal{C}=1$ QAH state is topologically equivalent to a $\mathcal{N}=2$ TSC.
Therefore, in the case of QAH$_{\mathcal{C}=1}$-TSC$_{\mathcal{N}=2}$-QAH$_{\mathcal{C}=1}$ junction (Fig.~\ref{fig3}j), the edge current will be perfectly transmitted. By contrast, if the junction is QAH$_{\mathcal{C}=1}$-TSC$_{\mathcal{N}=1}$-QAH$_{\mathcal{C}=1}$ (Fig.~\ref{fig3}i), the chiral edge state in the QAH region separates into two CMEMs at the TSC boundary~\cite{fu2009a,akhmerov2009}. One CMEM is perfectly transmitted, while the other is totally reflected. The edge transport of the junction is governed by the generalized Landauer-B\"{u}ttiker formalism, which includes the contributions from both the normal scattering and Andreev scattering~\cite{anantram1996,entin2008}. The general relationship between current and voltage on lead 1 and 2 shown in Fig.~\ref{fig1} is $I_1=(e^2/h)[(1-\mathcal{R}+\mathcal{R}_A)(V_1-V^0_{\mathrm{sc}})-(\mathcal{T}'-\mathcal{T}'_A)(V_2-V^0_{\mathrm{sc}})]$, and $I_2=(e^2/h)[(1-\mathcal{R}'+\mathcal{R}'_A)(V_2-V^0_{\mathrm{sc}})-(\mathcal{T}-\mathcal{T}_A)(V_1-V^0_{\mathrm{sc}})]$.
Here $V^0_{\mathrm{sc}}=0$ is the voltage of the grounded SC layer, $I_1$ and $I_2$ are currents flowing into leads 1 and 2, respectively. $\mathcal{R}$, $\mathcal{T}$, $\mathcal{R}_A$ and $\mathcal{T}_A$ are the normal reflection, normal transmission, Andreev reflection and Andreev transmission probabilities for an electron injected from the left, while $\mathcal{R}'$, $\mathcal{T}'$, $\mathcal{R}'_A$, and $\mathcal{T}'_A$ are for an electron coming from the right. The two-terminal conductance is then defined as $\sigma_{12}\equiv I/(V_1-V_2)=(I_1-I_2)/2(V_1-V_2)$.
For the QAH$_{\mathcal{C}=1}$-TSC$_{\mathcal{N}=1}$-QAH$_{\mathcal{C}=1}$ junction in Fig.~\ref{fig3}i, the probabilities of normal scattering and Andreev scattering are equal~\cite{chung2011}, and we have $\mathcal{R}=\mathcal{R}_A=\mathcal{T}=\mathcal{T}_A=\mathcal{R}'=\mathcal{R}'_A=\mathcal{T}'=\mathcal{T}'_A=1/4$, resulting in a half-quantized conductance
\begin{equation}
\sigma_{12}=\frac{e^2}{h}(\mathcal{T}+\mathcal{R}_A)=\frac{e^2}{2h}.
\end{equation}
Moreover, since the SC layer is not floating but grounded, the quantized net incoming current $I_{\text{SC}}=(V_1+V_2)e^2/h$ will be flowing from the SC layer to ground. Here we point out that the supercurrent due to the phase fluctuation of SC order parameter may give a small correction to conductance, which scales as $(\ell/L)^3$, where $\ell$ is the width of CMEM, and $L$ is the size of SC. For an estimation, $\ell\sim0.5~\mu$m, therefore such correction is neglible for $L>50~\mu$m. In contrast, the $\mathcal{N}=2$ TSC junction in Fig.~\ref{fig3}j exhibits a quantized conductance $\sigma_{12}=e^2/h$~\cite{chung2011}.

The entire plateau transition of $\sigma_{12}$ in the hybrid junction device is shown in Fig.~\ref{fig3}c. In correspondence to the QAH plateau transition of $\sigma_{xy}$ in Fig.~\ref{fig3}b, $\sigma_{12}$ also exhibits plateaus quantized at $e^2/h$ and $0$ when region II is $\mathcal{N}=\pm2$ TSC and $\mathcal{N}=0$ NSC, respectively. In addition, an intermediate half-quantized plateau at $e^2/2h$ could occur at the coercivity under the condition $|\Delta|+|m_0|>|\lambda|>|m_0|$, which is a unique signature of the $\mathcal{N}=\pm1$ TSC in region II. We emphasize that a plateau usually indicates a stable phase instead of a fine-tuned state. The size of backscattering region is not necessarily mesoscopic. In fact, the size $L$ of the TSC region sets a temperature scale $k_BT_{\text{int}}\sim v_M/L$, above which the interference effect vanishes due to thermal averaging, where $v_M$ is the Fermi velocity of CMEM. For an estimation, $L\sim200~\mu$m, $v_M\sim2.0$~eV \AA, $T_{\text{int}}\sim10$~mK. Therefore, the half-plateau is robust at large $L$ and finite temperature $T>T_{\text{int}}$. The plateau transitions and corresponding edge transport configuration in the hysteresis loop are illustrated in Fig.~\ref{fig3}c-j. In particular, \emph{four} $1/2$-plateaus occur around the critical magnetic fields $\pm H_1^*$ and $\pm H_2^*$ shown in Fig.~\ref{fig3}c.

\subsection{Point contact}

\begin{figure}[t]
\begin{center}
\includegraphics[width=3.3in]{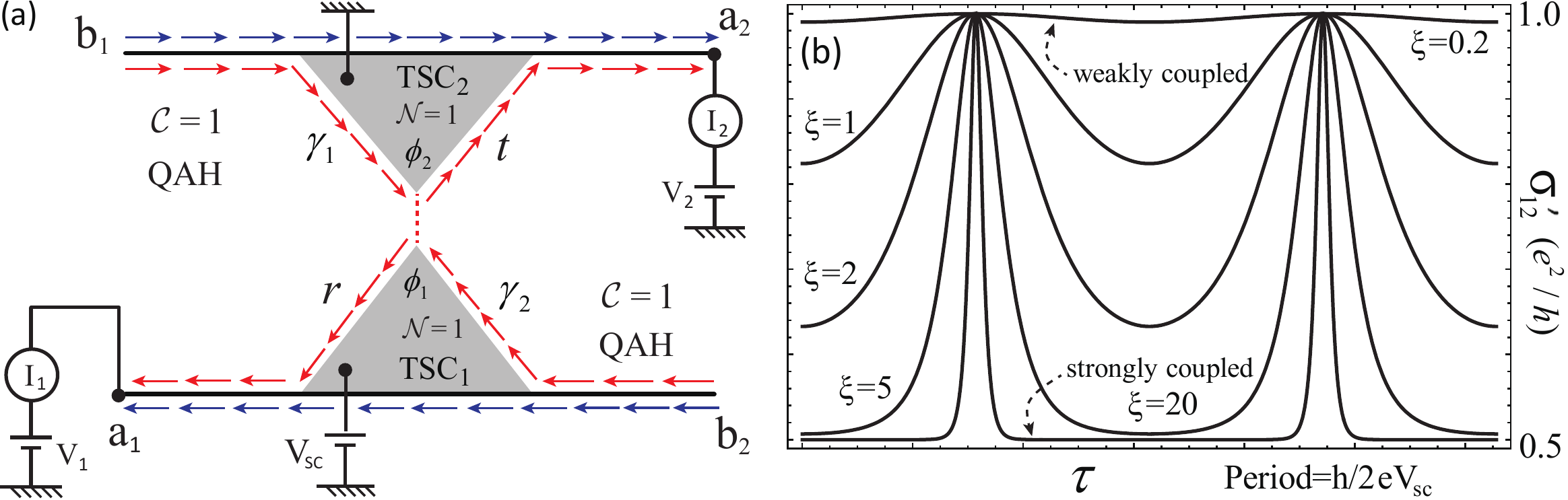}
\end{center}
\caption{(a) The point contact configuration of two SC islands with SC phases $\phi_1$ and $\phi_2$, across which the reflection and transmission amplitudes of the CMEMs are $r$ and $t$. (b) The conductance $\sigma_{12}'$ as a function of $\tau$ for different coupling strengths $\xi$. A dc current flows between $a_1$ and $a_2$, an ac voltage between them is measured, with frequency $f=2eV_{\text{sc}}/h$.
}
\label{fig4}
\end{figure}

Another useful transport configuration is a point contact formed by two SC islands which allow the transmission of CMEMs, as shown in Fig.~\ref{fig4}a. A voltage $V_{\mathrm{sc}}$ is applied onto island TSC$_1$, while TSC$_2$ is grounded. If either TSC$_1$ or TSC$_2$ is a $\mathcal{N}=2$ TSC, the edge current will be perfectly transmitted. Non-trivial physics occurs when both TSC$_1$ and TSC$_2$ are $\mathcal{N}=1$ TSC. An incident edge electron from $b_1$ splits into two CMEMs, one is perfectly transmitted along the edge, while the other is scattered at the point contact with transmission amplitude $t$, which depends on the phase difference $\delta\phi\equiv\phi_1-\phi_2$ of two TSCs (see Appendix B). The $I$-$V$ relation in this geometry is $I_1=(e^2/h)[(1-\mathcal{R}+\mathcal{R}_A)(V_1-V_{\mathrm{sc}})-(\mathcal{T}'-\mathcal{T}'_A)V_2]$, and $I_2=(e^2/h)[(1-\mathcal{R}'+\mathcal{R}'_A)V_2-(\mathcal{T}-\mathcal{T}_A)(V_1-V_{\mathrm{sc}})]$.
where $\mathcal{R}=\mathcal{R}_A=\mathcal{R}'=\mathcal{R}'_A=r^2/4$, $\mathcal{T}=\mathcal{T}'=(1+t)^2/4$, $\mathcal{T}_A=\mathcal{T}'_A=(1-t)^2/4$, $r$ is reflection amplitude and $r^2+t^2=1$. Therefore, $I=e^2(1+t)(V_1-V_2-V_{\mathrm{sc}})/2h$. Note that the current is proportional to the tunneling amplitude $t$, not the tunneling probability. If $V_{\mathrm{sc}}=0$, we have $\sigma_{12}=(1+t)e^2/2h$, which directly measures $t$ of the neutral CMEMs. A finite $V_{\text{sc}}$ leads to a time dependent $\delta\phi$, which in turn affects $t$. A simple tunneling model for the CMEM is (also see Appendix B)
\begin{equation}\label{tunnel_model}
H_{\text{tunnel}} = i\sigma_z\partial_x-\kappa(x)\sin(\delta\phi/2-\phi_0)\sigma_y,
\end{equation}
where $\kappa(x)$ is nonzero in a finite interval, and the basis is the CMEMs $(\gamma_1,\gamma_2)$ shown in Fig.~\ref{fig4}a. The transmission amplitude $t$ at zero-energy in this model is $t(\delta\phi)=1/\cosh[\xi\sin(\delta\phi/2-\phi_0)]$, where $\xi=\int dx\kappa(x)/2$. Within this model, $t$ is purely real. With a fixed $V_{\mathrm{sc}}$ across the point contact, $\delta\phi$ varies linearly with time $\tau$ with a slope $d\delta\phi/d\tau=2eV_{\mathrm{sc}}/\hbar$. We can define a new conductance
\begin{equation}
\sigma_{12}'\equiv \frac{I}{V_1-V_2-V_{\mathrm{sc}}}=\frac{e^2}{2h}\left[1+t(\delta\phi)\right],
\end{equation}
which is a periodic function in time with the Josephson junction frequency $f=2eV_{\mathrm{sc}}/h$. Fig.~\ref{fig4}b shows $\sigma_{12}'$ as a function of time for different values of $\xi$. The time oscillation shape of $\sigma_{12}'$ are different for a weakly coupled point contact (small $\xi$) and a strongly coupled one (large $\xi$). However, $\sigma_{12}'$ always oscillates between $e^2/2h$ and $e^2/h$, since there is always at least one perfectly transmitted CMEM, which is also a unique feature of the $\mathcal{N}=1$ TSC state.

\subsection{Temperature dependence}

We further consider the temperature dependence of the above CMEM transmission (see Appendix D).
It is straightforward to see by a dimensional counting that $t(\delta\phi)$ in the above free Majorana fermion model is marginal, therefore it remains constant at low temperature $T$. When the leading four-fermion interaction (irrelevant) is included, the tunneling amplitude acquires a weak temperature dependence. For $V_{\text{sc}}=0$, in this case $\sigma_{12}'=\sigma_{12}$, the renormalization group analysis gives a power-law correction $\delta t\sim-\lambda^2_{p}T^6$ to $t$, where $\lambda_p$ is the bare fermion interaction strength. The conductance $\sigma_{12}'\propto(1+t)$ will therefore decrease as $T$ increases. This perturbative result is no longer valid above a characteristic temperature of $T_c\sim \lambda_{p}^{-1/3}$, when the correction $\delta t$ is comparable to $t$. For higher temperature $T_c<T\ll|\Delta|$, $t$ will flow towards $0$, and the two TSC islands will behave like a single connected TSC analogous to that shown in Fig.~\ref{fig1}. In this regime, one can formulate a similar point-contact tunneling model between the left and right edges of the new TSC as in Eq.~(\ref{tunnel_model}), but with an additional vortex tunneling through the bulk TSC. At high temperature, the leading contribution to $t$ then comes from the vortex tunneling, which leads to $t\sim\lambda^2_{\sigma} T^{-7/4}$, where $\lambda_{\sigma}$ is the bare vortex tunneling strength. Therefore, the half-quantized plateau in $\sigma_{12}$ remains robust in the high temperature regime $T_c<T\ll |\Delta|$.

\section{Discussion and experimental realization}

Finally, we discuss the feasibility of our proposals. Experimentally, to observe the $\mathcal{N}=\pm1$ chiral TSC and all of the four half-quantized conductance plateaus, a good proximity effect between SC and magnetic TI is necessary. Moreover, the critical field $H^{\perp}_{c}$ of SC should be larger than the coercivity $H^*_{1,2}$ in magnetic TI. From Ref.~\onlinecite{fengy2015,kou2015}, the estimated $H_1^*\sim0.05$~T and $H_2^*\sim0.2$~T. The candidate SC materials are Nb and NbSe$_2$. The bulk Nb is a type I SC with $T_{\text{sc}}=9.6$~K and $H^{\perp}_c\sim0.2$~T, while a thin film Nb becomes a type II SC with upper critical field $H^{\perp}_{c2}\sim1$~T. NbSe$_2$ is a type II SC and shows good proximity effect with Bi$_2$Se$_3$~\cite{wangmx2012} even at $4.2$~K and $0.4$~T, where the proximity effect induced SC gap is $\Delta\sim0.5$~meV. The width of the CMEM $\ell$ can be estimated as $v_F/\Delta\sim0.52~\mu$m, where the Fermi velocity $v_F\sim2.6$~eV \AA~\cite{chang2013b}. For a typical junction voltage $V_{\text{sc}}\sim1~\mu$V, $f\sim0.48$~GHz, which is easily accessible in experiments.

\section{Conclusion}

In summary, we propose to realize the $\mathcal{N}=\pm1$ chiral TSC in a magnetic TI near the QAH plateau transition via the proximity effect to an $s$-wave SC. We show that inequivalent SC pairing amplitude on top and bottom surfaces in doped magnetic TIs will optimize the $\mathcal{N}=\pm1$ chiral TSC phases. Several edge transport measurements have been proposed to identify such $\mathcal{N}=1$ TSC in the QAH-SC hybrid system. In particular, the conductance could be quantized into a half-integer plateau at the coercive field in this hybrid system, as a unique signature of the neutral CMEM backscattering. We emphasize that such an experiment can work at reasonable temperature and does not depend on the interference effect of CMEM. We hope the theoretical work here can aid the search for chiral TSC phases in hybrid systems.

\begin{acknowledgments}
We thank David Goldhaber-Gordon and Andre Broido for useful comments on the draft. This work is supported by the US Department of Energy, Office of Basic Energy Sciences, Division of Materials Sciences and Engineering, under Contract No.~DE-AC02-76SF00515 and in part by FAME, one of six centers of STARnet, a Semiconductor Research Corporation program sponsored by MARCO and DARPA.
\end{acknowledgments}

\begin{appendix}

\section{Phase diagram under complex $\alpha=\Delta_2/\Delta_1$}
In the paper we have only considered the case $\alpha=\Delta_2/\Delta_1$ is real. In general, in the absence of time reversal symmetry (as is in our model), $\alpha=|\alpha|e^{i\phi_\alpha}$ is complex. Correspondingly, the phase diagram will be modified quantitatively, but the topology of the phase boundaries remains unchanged compared to those shown in Fig.~\ref{fig2} of the paper.

\begin{figure}[htbp]
\includegraphics[width=3.3in]{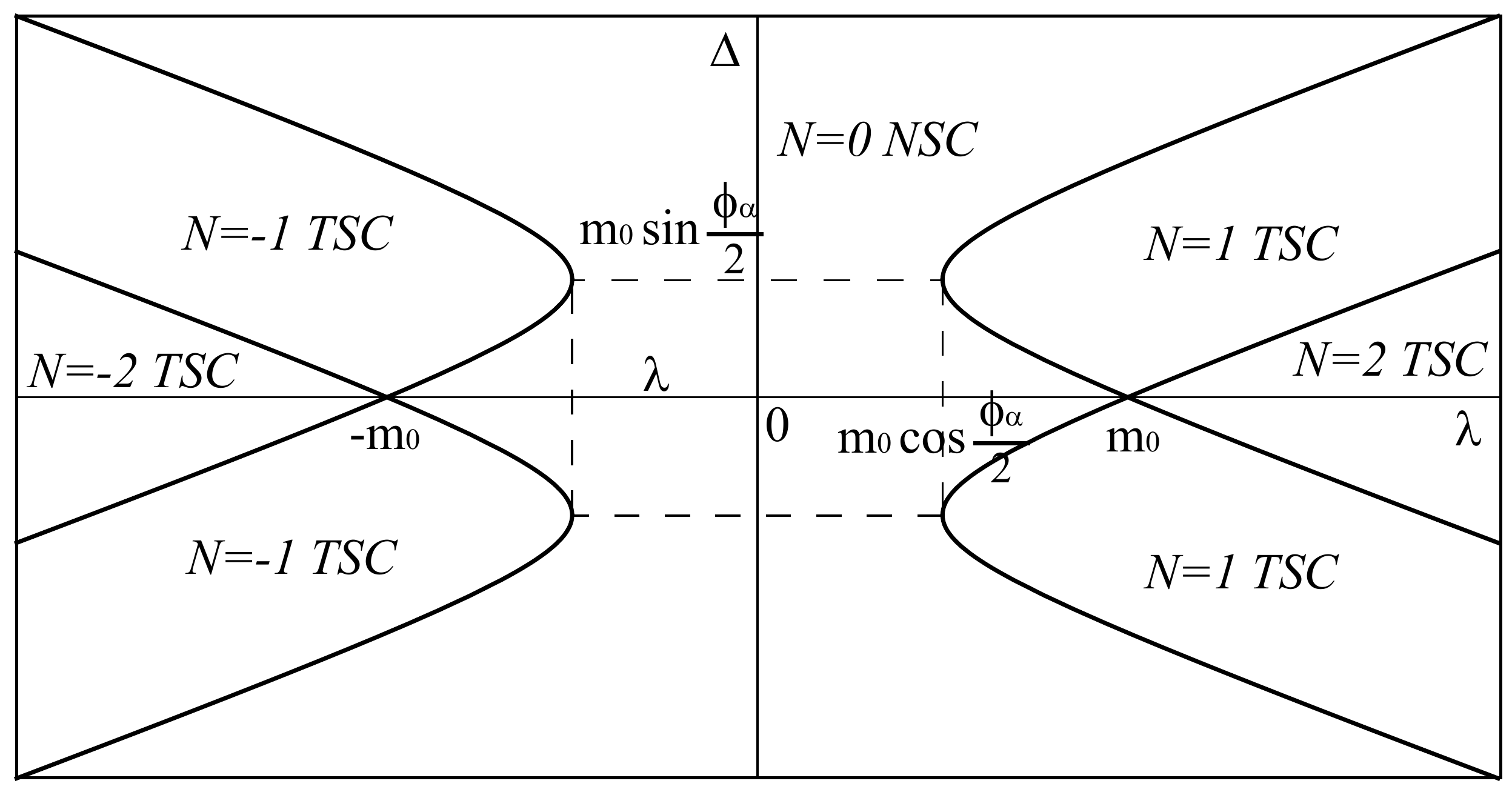}
\caption{The phase diagram for $\Delta_2=e^{i\phi_\alpha}\Delta_1$ and $\mu=0$. When $\phi_\alpha=0$, the $\mathcal{N}=\pm1$ TSC phases disappear, while when $\phi_\alpha=\pi$, the phase spaces of $\mathcal{N}=1$ and $\mathcal{N}=-1$ TSC touch each other, as indicated in Fig.~2b and Fig.~3a of the main text, respectively.}\label{fig5}
\end{figure}

By a proper choice of basis we can always set $\Delta_1=\Delta$ real. As an illustrative example, we consider here the case $|\alpha|=1$, namely $\alpha=\Delta_2/\Delta_1=e^{i\phi_\alpha}$. Via a unitary transformation $(c^t_{\mathbf{k}\uparrow}, c^t_{\mathbf{k}\downarrow}, c^b_{\mathbf{k}\uparrow}, c^b_{\mathbf{k}\downarrow})\rightarrow (c^t_{\mathbf{k}\uparrow}, c^t_{\mathbf{k}\downarrow}, e^{i\phi_\alpha/2}c^b_{\mathbf{k}\uparrow}, e^{i\phi_\alpha/2}c^b_{\mathbf{k}\downarrow})$, $\Delta_2$ is transformed into a real number $\Delta_2'=\Delta_1=\Delta$, while the hybridization $m(k)$ between the top and bottom SS becomes a complex number $e^{-i\phi_\alpha/2}m(k)$. Therefore, we can always set two of the three parameters $\Delta_1$, $\Delta_2$ and $m_0$ to real numbers. Diagonalizing the BdG Hamiltonian $H_{\text{BdG}}$ yields the energy spectrum $E^2=k^2+\Big[\lambda\pm\sqrt{\left[m(k)\sin(\phi_\alpha/2)\pm\Delta\right]^2+m(k)^2\cos^2\left(\phi_\alpha/2\right)}\Big]^2$.
The phase boundaries are given by the gap closing of the energy spectrum:
\begin{equation}
\lambda\pm\sqrt{(m_0\sin(\phi_\alpha/2)\pm\Delta)^2+m_0^2\cos^2\left(\phi_\alpha/2\right)}=0,
\end{equation}
namely, the following hyperbolas:
\begin{equation}
\lambda^2-\Big(\Delta\pm m_0\sin(\phi_\alpha/2)\Big)^2=m_0^2\cos^2\left(\phi_\alpha/2\right).
\end{equation}
The phase diagram is shown in Fig.~\ref{fig5}. As one can see, the topology of the phase diagram does not change much. In particular, when $\phi_\alpha=0$ and $\pi$, the phase diagram is as indicated in Fig.~2b and Fig.~3a of the paper, respectively.

\section{Derivation of the effective tunneling Hamiltonian}
Without loss of generality, consider the case $|\Delta|>\lambda-m_0>0$. The QAH has Chern number $\mathcal{C}=1$ and the SC in region II has BdG Chern number $\mathcal{N}=1$, both of which come from the lower block $H_{-}(\mathbf{k})$ of the BdG Hamiltonian $H_{\text{BdG}}$.
When the pairing amplitude of the superconductor is $\Delta=|\Delta|e^{i\phi}=\Delta_1=-\Delta_2$ with a phase $\phi$, $H_{-}(\mathbf{k})$ can be rewritten as
\begin{align}
H_-(\mathbf{k})&=
\begin{pmatrix}
h'_+(\mathbf{k}) & 0
\\
0 & -h'^*_-(-\mathbf{k})
\end{pmatrix},
\\
h'_\pm(\mathbf{k})&=
\begin{pmatrix}
m(k)-\lambda\pm|\Delta| & -ik_x \pm k_y
\\
ik_x\pm k_y & -m(k)+\lambda\mp|\Delta|
\end{pmatrix},
\end{align}
under the following new basis
$\frac{1}{\sqrt{2}}(e^{-i\phi/2}c_{\mathbf{k}\downarrow}+e^{i\phi/2}c^\dag_{-\mathbf{k}\uparrow}, e^{-i\phi/2}c_{\mathbf{k}\uparrow}+e^{i\phi/2}c^\dag_{-\mathbf{k}\downarrow}, -e^{-i\phi/2}c_{\mathbf{k}\downarrow}+e^{i\phi/2}c^\dag_{-\mathbf{k}\uparrow}, -e^{-i\phi/2}c_{\mathbf{k}\uparrow}+e^{i\phi/2}c^\dag_{-\mathbf{k}\downarrow})$,
where we have used the notation
\begin{align}
c_{\mathbf{k}\uparrow}=\frac{c^t_{\mathbf{k}\uparrow}-c^b_{\mathbf{k}\uparrow}}{\sqrt{2}},
\end{align}
and
\begin{align}
c_{\mathbf{k}\downarrow}=\frac{c^t_{\mathbf{k}\downarrow}+c^b_{\mathbf{k}\downarrow}}{\sqrt{2}}.
\end{align}
The Majorana edge state between the QAH (where $|\Delta|=0$) and the TSC (where $|\Delta|>\lambda-m_0>0$) is given by $h'_+(\mathbf{k})$.

As shown in Fig.~\ref{fig4} of the paper, the lower TSC$_1$ and the upper TSC$_2$ have superconducting phases $\phi_1$ and $\phi_2$ respectively. For simplicity, we shall approximate $m(k)$ as $m_0$, which does not change the topological physics. If the upper edge of the lower TSC$_1$ is set as $y=0$, the Hamiltonian of the corresponding Majorana edge state can be derived as
\begin{equation}
H_1=\int dx \ i\gamma_1(x)\partial_x\gamma_1(x),
\end{equation}
where
\begin{eqnarray}
\gamma_1(x) &=& \frac{e^{-i\phi_1/2}c_1(x)+e^{i\phi_1/2}c^\dag_1(x)}{\sqrt{2}},
\\
c_1(x) &=& \int_{-\infty}^{\infty}e^{(|\Delta|\Theta(-y)+m_0-\lambda)y}\left[e^{i\pi/4}c_{\uparrow}(x,y)\right.
\nonumber
\\
&&\left.+e^{-i\pi/4}c_{\downarrow}(x,y)\right]dy,
\end{eqnarray}
with $\Theta(y)$ defined as the Heaviside function.
Similarly, the lower edge of the upper TSC$_2$ at $y=y_0>0$ has a low energy Hamiltonian
\begin{equation}
H_2=-\int dx \ i\gamma_2(x)\partial_x\gamma_2(x),
\end{equation}
where
\begin{eqnarray}
\gamma_2(x) &=& \frac{e^{-i\phi_2/2}c_2(x)+e^{i\phi_2/2}c^\dag_2(x)}{\sqrt{2}},
\\
c_2(x) &=& \int_{-\infty}^{\infty}e^{(\lambda-m_0-|\Delta|\Theta(y-y_0))y}\left[e^{-i\pi/4}c_{\uparrow}(x,y)\right.
\nonumber
\\
&&\left.+e^{i\pi/4}c_{\downarrow}(x,y)\right]dy.
\end{eqnarray}
We shall assume the point contact extends in the interval $0<x<L$, and the two edges have a nonzero hopping and pairing term:
\begin{eqnarray}
H_{I} &=& -\int_0^L dx \left[J_hc_1^\dag(x)c_2(x)\right.
\nonumber
\\
&&\left.+J_p(\Delta_1^*+\Delta_2^*) c_1(x)c_2(x)+\text{h.c.}\right],
\end{eqnarray}
where $\Delta_{1,2}=|\Delta|e^{i\phi_{1,2}}$. When projected into the low energy Hilbert space of $\gamma_1$ and $\gamma_2$ via the substitutions
\begin{equation}
c_1\rightarrow e^{i\phi_1/2}\gamma_1/\sqrt{2}, \ \ c_2\rightarrow e^{i\phi_2/2}\gamma_2/\sqrt{2},
\end{equation}
this term becomes:
\begin{align}
H_{I} &= 2\int_0^L dx\ i\kappa(x)\sin\left(\frac{\delta\phi}{2}-\phi_0\right)\gamma_1(x)\gamma_2(x)
\nonumber
\\
&= 2\int_0^L dx\  i\lambda(x)\gamma_1(x)\gamma_2(x),
\end{align}
where
\begin{align}
\delta\phi &=\phi_1-\phi_2,
\\
\kappa(x) &=\left|J_h/2+i\text{Im}(J_p)\right|,
\\
\phi_0 &=\arg\left[J_h+i2\text{Im}(J_p)\right].
\end{align}
For simplicity we have defined
\begin{align}
\lambda(x)\equiv\kappa(x)\sin\left(\frac{\delta\phi}{2}-\phi_0\right).
\end{align}
The total tunneling Hamiltonian is then $H_{\text{tunnel}}=H_1+H_2+H_I$ as given in Eq.~(\ref{tunnel_model}) of the paper. The eigenwavefunction $\psi=(\eta_1,\eta_2)^T$ at energy $E$ can then be obtained by solving the following Shr\"{o}dinger equation:
\begin{equation}
\begin{pmatrix}
i\partial_x & i\lambda(x)
\\
-i\lambda(x) & -i\partial_x
\end{pmatrix}
\begin{pmatrix}
\eta_1
\\
\eta_2
\end{pmatrix}
=E\begin{pmatrix}
\eta_1
\\
\eta_2
\end{pmatrix}
\end{equation}
The solution for a wave incident from $x=-\infty$ with momentum $k$ is $E=k$, and
\begin{align}
&\left(\eta_1(x), \eta_2(x)\right)
\nonumber
\\
&=\left\{
\begin{array}{l@{\;\quad\;}l}
\left(e^{-ikx}, \frac{\lambda\sinh(\sqrt{\lambda^2-k^2}L)}{\mathcal{G}[L]}e^{ikx}\right) & (x\le0)\\
\left(\frac{\mathcal{G}[L-x]}{\mathcal{G}[L]}, \frac{\lambda\sinh[\sqrt{\lambda^2-k^2}(L-x)]}{\mathcal{G}[L]}\right) & (0<x\le L)\\
\left(\frac{\sqrt{\lambda^2-k^2}e^{-ikx}}{\mathcal{G}[L]},0\right) & (x>L)
\end{array}
\right.
\end{align}
where function $\mathcal{G}[x]=\sqrt{\lambda^2-k^2}\cosh(\sqrt{\lambda^2-k^2}x)-ik\sinh(\sqrt{\lambda^2-k^2}x)$. At low energies $k\ll\lambda$, the wavefunction can be approximately written as
\begin{align}
&\left(\eta_1(x),\ \eta_2(x)\right)
\nonumber
\\
&=\frac{1}{\cosh\lambda L}\left( \cosh\left[\int_x^\infty\lambda(x')dx'\right], \sinh\left[\int_x^\infty\lambda(x')dx'\right]\right),
\end{align}
from which the transmission and reflection amplitudes can be extracted out as
\begin{align}
t &= \frac{1}{\cosh\left(\int dx\lambda(x)\right)}=\frac{1}{\cosh\left[\xi\sin(\delta\phi/2-\phi_0)\right]},
\\
r &= \tanh\left(\int dx\lambda(x)\right)=\tanh \left[\xi\sin(\delta\phi/2-\phi_0)\right],
\end{align}
where $\xi=\int dx\kappa(x)$. Note that $t$ is always real and positive at low energies. For scattering at a finite energy $E=k$, the transmission amplitude $t$ is generally complex.

\section{S-matrix and conductance in general Josephson junction setup}

Here we formulate the scattering matrix of edge states in the setup of Fig.~4a, and derive the conductance $\sigma_{12}'$.
The edge fermions at four ends of the sample are denoted by $a_{1,2}$ and $b_{1,2}$ as shown in Fig.~4a. With transmission coefficient $t$ and reflection coefficient $r$ at the point contact, the scattering matrix $S$ due to the point contact is
\begin{align}
&\begin{pmatrix}
a_{1,\mathbf{k}}+a^{\dagger}_{1,-\mathbf{k}}
\\
a_{1,\mathbf{k}}-a^{\dagger}_{1,-\mathbf{k}}
\\
a_{2,\mathbf{k}}+a^{\dagger}_{2,-\mathbf{k}}
\\
a_{2,\mathbf{k}}-a^{\dagger}_{2,-\mathbf{k}}
\end{pmatrix}
=S\begin{pmatrix}
b_{1,\mathbf{k}}+b^{\dagger}_{1,-\mathbf{k}}
\\
b_{1,\mathbf{k}}-b^{\dagger}_{1,-\mathbf{k}}
\\
b_{2,\mathbf{k}}+b^{\dagger}_{2,-\mathbf{k}}
\\
b_{2,\mathbf{k}}-b^{\dagger}_{2,-\mathbf{k}}
\end{pmatrix}
\nonumber
\\
&=\begin{pmatrix}
r & 0 & t & 0\\
0 & 0 & 0 & 1\\
t^* & 0 & -r^* & 0\\
0 & 1 & 0 & 0
\end{pmatrix}
\begin{pmatrix}
b_{1,\mathbf{k}}+b^{\dagger}_{1,-\mathbf{k}}
\\
b_{1,\mathbf{k}}-b^{\dagger}_{1,-\mathbf{k}}
\\
b_{2,\mathbf{k}}+b^{\dagger}_{2,-\mathbf{k}}
\\
b_{2,\mathbf{k}}-b^{\dagger}_{2,-\mathbf{k}}
\end{pmatrix}.
\end{align}
Upon basis transformation from Majorana fermions to charged fermions on QAH edges, we have
\begin{align}
\begin{pmatrix}
a_{1,\mathbf{k}}
\\
a^{\dagger}_{1,-\mathbf{k}}
\\
a_{2,\mathbf{k}}
\\
a^{\dagger}_{{2,-\mathbf{k}}}
\end{pmatrix}
&=
\frac{1}{2}\begin{pmatrix}
r & r & t+1 & t-1\\
r & r & t-1 & t+1\\
t^*+1 & t^*-1 & -r^* & -r^*\\
t^*-1 & t^*+1 & -r^* & -r^*
\end{pmatrix}
\begin{pmatrix}
b_{1,\mathbf{k}}
\\
b^{\dagger}_{1,-\mathbf{k}}
\\
b_{2,\mathbf{k}}
\\
b^{\dagger}_{2,-\mathbf{k}}
\end{pmatrix},
\end{align}
based on which the normal/Andreev transmission/reflection probabilities are given as $\mathcal{T}=|t+1|^2/4$,
$\mathcal{T}_A=|t-1|^2/4$, and $\mathcal{R}=\mathcal{R}_A=|r|^2/4$.
According to the generalized Landauer-B\"{u}ttiker formula, the conductance defined in the main text is
\begin{equation}
\sigma_{12}'=\frac{1+\mathrm{Re}(t)}{2}\frac{e^2}{h}.
\end{equation}
Note that the conductance $\sigma_{12}'$ merely depends on the real part of Majorana transmission coefficient $t$, physically it is due to the fact that charged fermions are treated as combinations of Majorana fermions with transmissions $t$ and perfect transmission $1$.

\section{Temperature dependence and renormalization group analysis}
In this section we analyze the temperature dependence of Majorana transmission coefficient $t$ by renormalization group technique in detail~\cite{kane1992,kane2007}. Specifically, we focus on its real part $\mathrm{Re}(t)$, since it is directly related to the conductance $\sigma_{12}$. Our starting point is the action for the model in Eq.~(\ref{tunnel_model}) of the paper,
\begin{align}
\mathcal{S}_0=&\int d\tau \int dx[\gamma_1 i(\partial_{\tau}+\partial_x) \gamma_1
                    +\gamma_2 i(\partial_{\tau}-\partial_x) \gamma_2
                    \nonumber
                    \\
                    &+2\xi\delta(x)\sin(\delta\phi/2-\phi_0) i\gamma_1\gamma_2].
\end{align}
Since the Majorana tunneling occurs locally at $x=0$, the scaling dimension of the tunneling strength $\xi$ vanishes, i.e. $[\xi]=0$. Therefore, $\xi$ is invariant when the temperature $T$ of the system changes, and so does the transmission coefficient $t$.

The temperature dependence of $t$ comes from higher irrelevant terms at the point contact.
The leading irrelevant term is a four fermion interaction of the following form:
\begin{equation}
H_p=\int dx\lambda_{p}\delta(x)\gamma_1\partial_x\gamma_1\gamma_2\partial_x\gamma_2.
\end{equation}
It represents the tunneling of one pair of Majorana fermions from one edge to the other. The scaling dimension of $\lambda_p$ is $[\lambda_p]=-3$, hence it is irrelevant and scales as $\lambda_p^{\text{eff}}\sim\lambda_p T^3$ when $T\rightarrow 0$. Increasing the temperature $T$ will enhance the effective interaction strength $\lambda_p$, which affects the transmission coefficient $t$.

The contribution of $H_p$ to the transmission coefficient $t$ can be calculated perturbatively as follows. Suppose both $\xi$ and $\lambda_p$ are small, so that perturbation theory can be used. We shall regard $H_I=2i\xi\delta(x)\sin(\delta\phi/2-\phi_0)\gamma_1\gamma_2$ and $H_p$ given above as the  perturbation.
Consider an in-state $|i\rangle=\gamma_{1,-k}|\Omega\rangle$ of Majorana fermion $\gamma_1$, and a transmitted out-state $|f\rangle=\gamma_{1,-k'}|\Omega\rangle$, where $|\Omega\rangle$ is the system ground state. The transmission coefficient $t$ is then given by
\begin{equation}
t\approx\left\langle f\right|T_\tau e^{-i\int_{-\infty}^\infty (H_I+H_p) d\tau}\left|i\right\rangle,
\end{equation}
where $T_\tau$ stands for the time ordering. The zero-order $t^{(0)}$ is simply $\delta_{kk'}$. The first-order contribution $t^{(1)}$ is
\begin{equation}
t^{(1)}= \left\langle f\right|-i\int (H_{I}+H_p)d\tau\left|i\right\rangle.
\end{equation}
Since $H_I$ is odd in $\gamma_1$ and $\gamma_2$, its first-order contribution vanishes. The second term of $H_p$ is purely imaginary and therefore does not contribute to the conductance $\sigma_{12}'$. The second-order correction
\begin{equation}
t^{(2)}\sim -\frac{1}{2}\left\langle f\right|T_\tau\int(H_I+H_p)(\tau)(H_I+H_p)(\tau')d\tau d\tau'\left|i\right\rangle.
\end{equation}
The $H_I^2$ term gives a constant contribution $\sim -\xi^2\sin^2(\delta\phi/2-\phi_0)$, in agreement with calculations in Appendix B. The cross term $H_IH_p$ vanishes because it is odd in $\gamma_1$ and $\gamma_2$. The $H_p^2$ term results in a temperature dependent correction to the real part of transmission coefficient $t$ as
\begin{equation}
\delta\text{Re}(t)\sim-\delta_{kk'}\left(\lambda_{p}^{\text{eff}}\right)^2=-\delta_{kk'}\lambda_p^2 T^6.
\end{equation}
Therefore, the transmission coefficient $t$ generically decreases as temperature $T$ increases. When the temperature $T$ is above a characteristic temperature $T_c\sim\lambda_{p}^{-1/3}$, the interaction $\lambda_p$ at the point contact dominates, so that $t$ becomes small and  $r$ becomes large. In this case, the above perturbative treatment is no longer valid.
However, this case can be effectively viewed as a breaking up of original Majorana edge states $\gamma_1$ and $\gamma_2$ and a remerge of them into two new Majorana edge states $\psi_1$ and $\psi_2$ on the left and right of the point contact, and of the two TSCs merging into a single TSC. In the temperature range $T_c\ll T\ll |\Delta|$, we can do a perturbation calculation about the high temperature fixed point before the superconducting phase is destroyed.

This scenario is very similar with our setup in Fig.~1a, except that the two edges are brought together at the point contact. Since the region between the edges in this case is a SC, there are both fermion tunnelings and vortex tunnelings between edges~\cite{fendly2007}. The effective action for this point contact is
\begin{align}
\mathcal{S}'=& \int d\tau \int dy\left[\psi_1 i(\partial_{\tau}+v_m \partial_y) \psi_1
                    +\psi_2 i(\partial_{\tau}-v_m \partial_y) \psi_2\right.
                    \nonumber
                    \\
                    &\left.+\lambda_\psi \delta(y) i\psi_1\psi_2
                    +\lambda_\sigma \delta(y) \sigma_1\sigma_2\right],
\end{align}
where $\sigma_1$ and $\sigma_2$ are the vortex operators on edges with a scaling dimension $[\sigma_1]=[\sigma_2]=1/16$. Dimension counting renders $[\lambda_\psi]=0$ and $[\lambda_\sigma]=7/8$, so the vortex-vortex tunneling is the most relevant. Therefore, at a high temperature $T$, the vortex-vortex tunneling term gives the temperature dependence of transmission coefficient $t$
\begin{equation}
t\sim\lambda^{2}_\sigma T^{-7/4}.
\end{equation}
The power-law relation is valid above a characteristic temperature $T'_c\sim \lambda^{8/7}_\sigma$, provided the SC gap $|\Delta|$ is much higher. In fact, this confirms the robustness of the half-quantized plateau. For in the setup with reasonable finite temperature, the edges are far away from each other, so the tunneling strengths including $\lambda_{\sigma}$ are sufficiently tiny, resulting in an extremely low $T_c'$.

\end{appendix}


\begin{thebibliography}{52}
\bibitem [{bas()}]{basis_note}%
  \BibitemOpen
  \href@noop {} {}\bibinfo {note} {The basis in Eq.~(\ref{decoupled}) is
  $(c_{t\mathbf{k}\uparrow}+c_{b\mathbf{k}\uparrow},c_{t\mathbf{k}\downarrow}-c_{b\mathbf{k}\downarrow},
  c^{\dag}_{t\overline{\mathbf{k}}\uparrow}+c^{\dag}_{b\overline{\mathbf{k}}\uparrow},
  c^{\dag}_{t\overline{\mathbf{k}}\downarrow}-c^{\dag}_{b\overline{\mathbf{k}}\downarrow})^T/\sqrt{2}$
  for $H_+(\mathbf{k})$ and
  $(c_{t\mathbf{k}\downarrow}+c_{b,\mathbf{k}\downarrow},c_{t\mathbf{k}\uparrow}-c_{b\mathbf{k}\uparrow},
  c^{\dag}_{t\overline{\mathbf{k}}\downarrow}+c^{\dag}_{b\overline{\mathbf{k}}\downarrow},
  c^{\dag}_{t\overline{\mathbf{k}}\uparrow}-c^{\dag}_{b\overline{\mathbf{k}}\uparrow})^T/\sqrt{2}$
  for $H_-(\mathbf{k})$. $\overline{\mathbf{k}}\equiv-\mathbf{k}$.}\BibitemShut
  {Stop}%
\end{thebibliography}
\end{document}